# A 2D van der Waals Material for Terahertz Emission with Giant Optical Rectification

Taketo Handa[1], Chun-Ying Huang[1], Yiliu Li[1], Nicholas Olsen[1], Daniel G. Chica[1], David D. Xu[1], Felix Sturm[2], James W. McIver[2], Xavier Roy[1], and Xiaoyang Zhu[1,*]

1. Department of Chemistry, Columbia University, New York, NY 10027, USA
2. Department of Physics, Columbia University, New York, NY 10027, USA

**ABSTRACT. Exfoliation and stacking of two-dimensional (2D) van der Waals (vdW) crystals have created unprecedented opportunities in the discovery of quantum phases. A major obstacle to the advancement of this field is the limited spectroscopic access due to a mismatch in sample sizes ($10^{-6} - 10^{-5}$ m) and wavelengths ($10^{-4} - 10^{-3}$ m) of electromagnetic radiation relevant to their low-energy excitations. Here, we introduce a new member of the 2D vdW material family: a terahertz (THz) emitter. We show intense and broadband THz generation from the vdW ferroelectric semiconductor $NbOI_2$ with optical rectification efficiency over one-order-of-magnitude higher than that of the current standard THz emitter, ZnTe. The $NbOI_2$ THz emitter can be easily integrated into vdW heterostructures for on-chip near-field THz spectroscopy of a target vdW material/device. Our approach provides a general spectroscopic tool for the rapidly expanding field of 2D vdW materials and quantum matter.**

Stacking two-dimensional (2D) materials into artificial van der Waals (vdW) structures has enabled the realization of a rich array of physical properties, particularly various quantum phases of matter[1]. Successful experimental demonstrations to date include, among many others, correlated insulators[2–5], superconductors[6,7], and fractional quantum anomalous Hall effects[8–12]. These discoveries highlight the importance of accessing the low-energy electronic landscapes, such as flat bands and correlated energy gaps, in understanding 2D vdW quantum matter. For the majority of quantum phases discovered, the relevant energy scales are estimated in the range of a few to a few tens of meV, i.e., the terahertz (THz) frequency range, as suggested by theoretical calculations[13–16] and experimental measurements of critical temperatures[2–12]. While scanning probe spectroscopy techniques have been able to determine the low-energy moiré flat bands[17–20], they require spatial access to the active material, making them difficult to implement in commonly used device architectures involving hexagonal boron nitride (*h*-BN) encapsulation and top/bottom gates. An ideal approach would be direct optical access to the 2D quantum matter in the relevant energy range, but doing so in the THz window has been challenging due to a mismatch in the lateral dimensions of 2D vdW devices ($10^{-6} - 10^{-5}$ m) and the wavelength ($10^{-4} - 10^{-3}$ m) of THz radiation. Although an infrared (IR) photocurrent technique has been demonstrated for a graphene moiré device at the high energy end (> 10 meV)[21], this approach requires very high-quality ohmic contacts and is difficult for semiconductors, e.g., the widely used transition metal dichalcogenides[3–5,8–12]. While near-field approaches based on local tip-scattering[22] or on-chip THz waveguides[23–26] have been demonstrated for µm-scale 2D samples, extending these approaches to complex vdW device structures remains challenging.

---

[*] Corresponding author. Email: xyzhu@columbia.edu

Here, we report that a 2D vdW ferroelectric semiconductor $NbOI_2$ is a broadband and intense THz emitter due to giant optical rectification. We compare THz generation from $NbOI_2$ to that from ZnTe single crystal, the benchmark three-dimensional (3D) material in use today [27,28]. We show that the optical rectification efficiency of $NbOI_2$ is one to two orders-of-magnitude higher than that of ZnTe. Due to its vdW structure, $NbOI_2$ crystals can be easily exfoliated and the resulting flakes stacked into vdW structures of interest (Fig. 1a). The effective size of the THz wave at the target vdW material/structure is determined by the size of the diffraction-limited optical pump laser spot (~µm) in the near IR-visible spectral range. Hence, it provides a vdW assembly approach for near-field THz spectroscopy. This simple approach could be applicable to a broad range of 2D materials, vdW heterostructures, and the *in situ* and near-field implementation of THz time-domain spectroscopy (THz-TDS)[29–32] will provide direct spectroscopic insight into the low-energy landscape and complex conductivity spectra of 2D quantum phases and collective excitations.

Efficient and broadband THz emission

An ultrathin 2D THz emitter can be readily obtained from mechanical exfoliation of a bulk

NbOI$_2$ single crystal (Fig. 1a), with flake thicknesses ranging from nm to µm (Methods and Supplementary Fig. S1). For thinner flakes, the conventional pickup and transfer process with h-BN can be used (Supplementary Fig. S2). Each NbOI$_2$ monolayer possesses in-plane ferroelectricity due to the displacement of Nb and O ions along the crystallographic b-axis[33–35] (Fig. 1b), giving rise to a large second-order nonlinear susceptibility, $\chi^{(2)}$[34]. The large $\chi^{(2)}$ from each layer and the in-phase alignment of the symmetry-breaking electric polarizations from different layers (Fig. 1b) enable efficient optical rectification[36] for the generation of THz radiation propagating perpendicular to the 2D planes. This is in contrast to other vdW in-plane ferroelectrics[37] and transition metal dichalcogenides[38] in which the nonlinear responses from adjacent layers cancel out. The THz emission results reported below are obtained under normal incidence with femtosecond laser pulses at a center wavelength of 800 nm, which is below the bandgap of NbOI$_2$ (see Supplementary Fig. S3)[34,39]. The absence of photocarrier generation avoids heating or photobleaching. This critical property distinguishes NbOI$_2$ from other thin THz emitters based on photocarrier generation[40–42], including spintronic effects[43], where the accompanying heat generation[44] poses a formidable challenge for low-temperature applications.

The first favorable property of NbOI$_2$ as a vdW THz emitter is the record-breaking efficiency of optical rectification. Figure 1c shows the time-domain waveform of the emitted radiation from a 2.5 µm thick NbOI$_2$ flake (red), in comparison to that from a standard ZnTe THz emitter with 200 µm thickness (blue). Both samples are excited under the same conditions and the THz emission is detected at far-field by electro-optical (EO) sampling in a 1-mm ZnTe (Fig. 1c) or a 0.2-mm GaP (Supplementary Fig. S4) crystal. Despite being 80 times thinner, the NbOI$_2$ flake exhibits comparable THz field amplitude to that from ZnTe[27]. The waveform from NbOI$_2$ during the initial few cycles is nearly identical to that from ZnTe, suggesting that both result from the same mechanism, i.e. optical rectification[27]. After the few cycles, there is an additional but weaker ringing signal observed for NbOI$_2$, but not for ZnTe; the origin of this ringing signal will be discussed later. Figure 1d presents the frequency-domain spectra for NbOI$_2$ (red) and ZnTe (blue), obtained by Fourier-transforming the time-domain waveforms obtained with the GaP or the ZnTe (insets) EO detectors. The frequency cutoff for THz emission in ZnTe occurs at around 3.5 THz due to poor phase matching, while that from NbOI$_2$ shows a broader bandwidth up to 5.5 THz (23 meV) as phase matching requirement in the thin flake is relaxed.

The second favorable property of NbOI$_2$ as a vdW THz emitter is the linear power dependence over an exceptionally broad window. Figure 1e shows the pump fluence dependence of the peak THz field from the 2.5 µm thick $NbOI_2$ flake (red) and the 200 µm thick ZnTe crystal (blue). THz emission from $NbOI_2$ shows linearity in a broad pump fluence range (ρ), up to 10 mJ/cm². The linearity confirms optical rectification as the origin of broadband THz generation[36] (see Supplementary Information), with little side effects from linear or nonlinear absorption. A small deviation from the linear relationship (dashed line) is only observed for ρ > 10 mJ/cm². For comparison, THz generation from ZnTe (blue dots in Fig. 1e) shows a clear saturation behavior at ρ > 1 mJ/cm² due to the competing two-photon absorption process[45]. Similar saturation behavior has also been reported in other technologies due to either linear or nonlinear photoabsorption[40,46].

Because of the excellent linearity from $NbOI_2$, the absolute field strength from a 2.5 µm thick $NbOI_2$ flake is comparable to that from a 200 µm thick ZnTe at high pump fluences (inset in Fig. 1e), with peak electric field exceeding 300 V/cm. We speculate that the high saturation threshold of $NbOI_2$ arises from a selective enhancement of $\chi^{(2)}$ in $NbOI_2$ (Supplementary Information). The consistent performance under high fluence attests to the superior performance of $NbOI_2$ as a near-field local THz source pumped by diffraction limited beams in the near-IR to visible range. Under the pump fluence of several mJ/cm² in dry air, neither the THz peak intensity nor the waveform show any change in the course of days (Supplementary Fig. S6). The linearity further extends to the low fluence limit; this enables near-field THz spectroscopy even with pump pulse energy as low as a few pico-joules (pJ), as shown below.

Thickness-dependent THz emission

Figure 2a compares the THz waveforms (normalized) from $NbOI_2$ flakes with thickness $d$ = 2.5 µm (blue), 650 µm (red), and 31 nm (green). While the three flakes show the same waveform for the initial few cycles attributed to optical rectification, we observe that the relative amplitude of the ringing signal at longer delay times increases considerably as the thickness decreases. This observation suggests that, besides thickness-independent optical rectification mechanism, there is an additional mechanism for THz generation for the thinner flakes. This thickness dependence is clearly shown in the Fourier-transform spectra (lower panel in Fig. 2b). While the broad THz emission spectra from optical rectification are nearly identical for the three thicknesses, a narrow THz peak (full-width-at-half-maximum, FWHM = 0.08 THz) at 3.13 ± 0.06 THz dominates the emission spectrum from the thinner flakes. As detailed in supplementary text, we attribute this narrow THz peak to a mechanism in which the below-gap pulse launches coherent transverse optical (TO) phonon wavepacket via impulsive Raman scattering, induces a temporal variation of macroscopic ferroelectric polarization, and serves as a (ferro)electric dipole radiation source. The TO nature of the coherent phonon mode is confirmed in the THz transmission spectrum obtained by THz-TDS (upper panel in Fig. 2b), which reveals a sharp absorption peak at $3.130 \pm 0.008$ THz.

Figure 2c shows log-log plots of the thickness dependences of the frequency-integrated Fourier amplitudes ($I_{THz}$) for the broadband emission from optical rectification (green circles) and the narrow band emission at 3.13 THz (purple triangles). We observe distinct THz emission down to the 31 nm thick $NbOI_2$ flake (see also Fig. 2a); giving the excellent signal-to-noise ratio, we expect similar results for even thinner flakes. The solid line is fit to $I_{THz} \propto d^a$, where $a$ is an exponent. We obtained $a = 0.8 \pm 0.1$ and thus the relationship is close to linear, expected for optical rectification. The small deviation from the linear relationship for larger $d$ may be attributed to phase mismatch. For the narrow band emission, the THz field remains nearly constant for thicknesses larger than 400 nm likely due to strong re-absorption by the TO phonon mode. While the broadband spectrum from optical rectification can serve as a general radiation source for *in situ* and near-field THz-TDS to probe the low-energy excitations and complex conductivity spectra of vdW quantum matter, the narrow band THz radiation can be used as a monochromatic THz field, particular at the lower thickness limit, to coherently modulate a range of physical properties in the vdW heterostructures[47,48]. Adding to the favorable properties of $NbOI_2$ as a vdW THz emitter, we note that THz emission is nearly temperature-independent (Fig. 2d), presumably due to its very high ferroelectric-to-paraelectric transition temperature ($> 573$ K)[39].

vdW THz emitter allows near-field and *in situ* THz spectroscopy

To demonstrate the application of $NbOI_2$ for *in situ* near-field THz-TDS, we fabricate a vdW heterostructure consisting of 30 nm *h*-BN/60 nm graphite/30 nm *h*-BN on a 1.3 μm thick $NbOI_2$ flake (Fig. 3a). The lateral dimension of the target graphite is 25 μm, which is approximately one-order of magnitude smaller than the far-field THz wavelength. We chose graphite here since it shows broad absorption in the THz region due to free carriers[49]. To probe this broadband absorption, we use a relatively thick $NbOI_2$ flake where the broadband optical rectification dominates THz generation (see Fig. 2). Importantly, the vdW stack can be observed optically from the $NbOI_2$ side due to its transparency in the visible to near-IR (see Supplementary Fig. S9), enabling measurements in a simple transmission geometry (Fig. 3a). Figure 3b shows the local transmission of the peak THz field obtained by scanning the lateral position of the sample, with the pump pulse energy at 40 pJ and a spot diameter of 4.6 μm (lower panel in Fig. 3c). We observe a ~80% reduction of the THz transmission when graphite is behind the emitter and this reduction is attributed to free carrier absorption. We estimate a spatial resolution of $5.0 \pm 0.8$ μm for the THz field as 1/e diameter (upper panel in Fig. 3c); the corresponding spatial resolution in THz power is $2.5 \pm 0.4$ μm. Thus, the approach described here provides a sub-diffraction (~$10^{-2}\lambda$) and *in-situ* vdW THz source, without waveform distortion encountered in tip- or aperture-based approaches.

One critical advantage is that this approach readily allows the acquisition of a local time-domain waveform on the sample area and reference area (Fig. 3d). The observed reduction of the field amplitude and a slight phase shift are reminiscent of a broadband Drude-like absorption [29]. Using the time-domain data, we obtain the frequency-domain complex conductivity (Fig. 3e). Because the local emitter and sample thicknesses and their separation are all much smaller than the far-field THz wavelength, the THz wave detected at the far-field EO crystal is the convolution of the reference field and in-plane complex conductivity. We fit the complex spectrum to the Drude model $\sigma_{THz} = \sigma_{DC} / (1 - i\omega\tau)$, and obtaine the Drude parameters: $\tau = 50 \pm 6$ fs and $\sigma_{DC} = 32 \pm 3.0$ mS, where $\sigma_{DC}$ is the Drude weight, $\omega$ angular frequency, $\tau$ scattering time. Using the scattering time and the effective mass of graphite $m_{eff} = 0.03 m_e$ where $m_e$ is the free electron mass[50], we estimate a carrier mobility in the thin graphite flake of $3.0 \times 10^3$ cm$^2$V$^{-1}$s$^{-1}$, which is ~3x lower than the bulk mobility of $1.0 \times 10^4$ cm$^2$V$^{-1}$s$^{-1}$ at room temperature[50]. A reduction in mobility from bulk graphite is known when the thickness is decreased below ~100 nm[51,52].

Discussion

We have demonstrated highly efficient, broadband, and robust THz emission from the 2D vdW ferroelectric NbOI$_2$. The giant optical rectification efficiency from NbOI$_2$ is over one order of magnitude higher than that from the benchmark 3D material, ZnTe single crystal. By integrating this THz emitter into vdW heterostructures, we demonstrate *in situ* THz-TDS and microscopy with lateral resolution almost two orders of magnitude below the diffraction limit. The present approach brings the relevant features of THz-TDS to the μm length scale, which is suitable for probing quantum matter in complex vdW heterostructures demonstrated in recent experiments. In addition to the simultaneous determination of real and imaginary parts of THz conductivity spectra, in-plane symmetry of quantum phases could be probed by exploiting the polarization of the THz field, tunable by the incident pump polarization. The use of $NbOI_2$ with the broadband THz emission covering up to 5.5 THz (Fig. 1d), allows access to much higher frequencies than those accessible with other near-field techniques, e.g., on-chip THz waveguide which typically ranges from 20 GHz to 1 THz[26]. Furthermore, this approach can be easily adapted in an optical pump and THz probe scheme to interrogate the dynamics and potential hidden quantum phases in the vdW systems[53,54]. We note that the residual of the near-IR pump used for THz generation is expected to have minimal effect on the quantum phases, as recent experiments on Mott states in the $WS_2$/$WSe_2$ moiré system exhibit little change for pulse energy in the 10 pJ range[53,54]. When necessary, photonic layers may also be incorporated to reduce the transmission of near-IR or visible laser light into the target vdW structures. Our simple approach leverages the fundamental benefits of vdW assembly and fills a critical technology gap of low-energy optical probes of a wide range of 2D materials or vdW heterostructures.

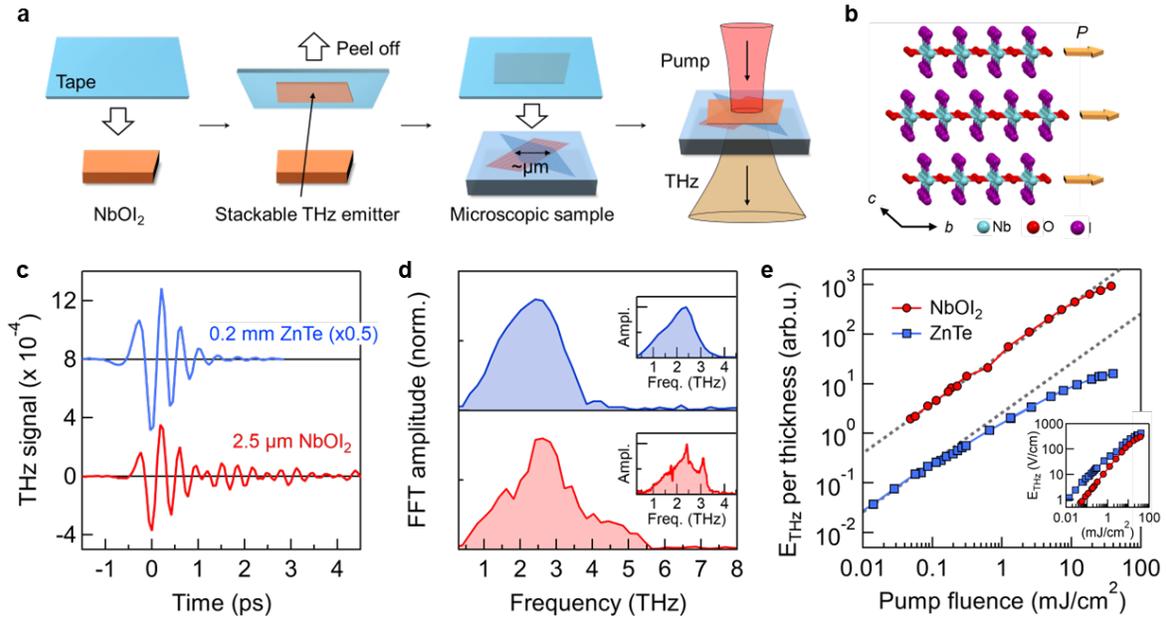

**Fig. 1 | Schematic illustration of vdW near-field THz spectroscopy that exploits highly efficient THz emission of 2D ferroelectric NbOI$_2$.** (a) A flat and ultrathin flake of NbOI$_2$ can be exfoliated using conventional tape exfoliation methods. This exfoliated flake is a readily stackable 2D THz emitter and is transferred onto a microscopic sample of interest. THz radiation is generated locally by a focused near-infrared pump. The spatial size of THz light at the sample is determined by the pump beam size. Thus, the transmitted THz wave contains the local conductivity information of the microscopic sample. (b) Crystal structure of NbOI$_2$. Spontaneous polarization (orange arrows) and accompanying inversion symmetry breaking exist along the *b*-axis. The direction of the spontaneous polarization does not flip along the stacking direction. (c) Time-domain THz waveform of 2.5-μm NbOI$_2$ (red) and 0.2-mm ZnTe (blue) under the excitation fluence of 1.24 mJ/cm$^2$ (86 nJ and 1/e radius of 47 μm derived from regenerative amplifier), recorded using 1-mm ZnTe as an EO detector. The emission signal for 0.2-mm ZnTe is multiplied by 0.5. (d) Frequency-domain spectra for NbOI$_2$ (red) and ZnTe (blue) recorded using a 0.2 mm thick GaP EO detector, which has a bandwidth up to 8 THz. The insets show the same measurements recorded using a 1 mm thick ZnTe EO detector. The main panels reveal the actual cutoff of the emitters, while the insets reveal the fine structure in the frequency domain. (e) Pump fluence dependence of the emitted field amplitudes, normalized by the thickness. The inset shows the absolute peak THz amplitude, estimated based on the equation for the EO effect.

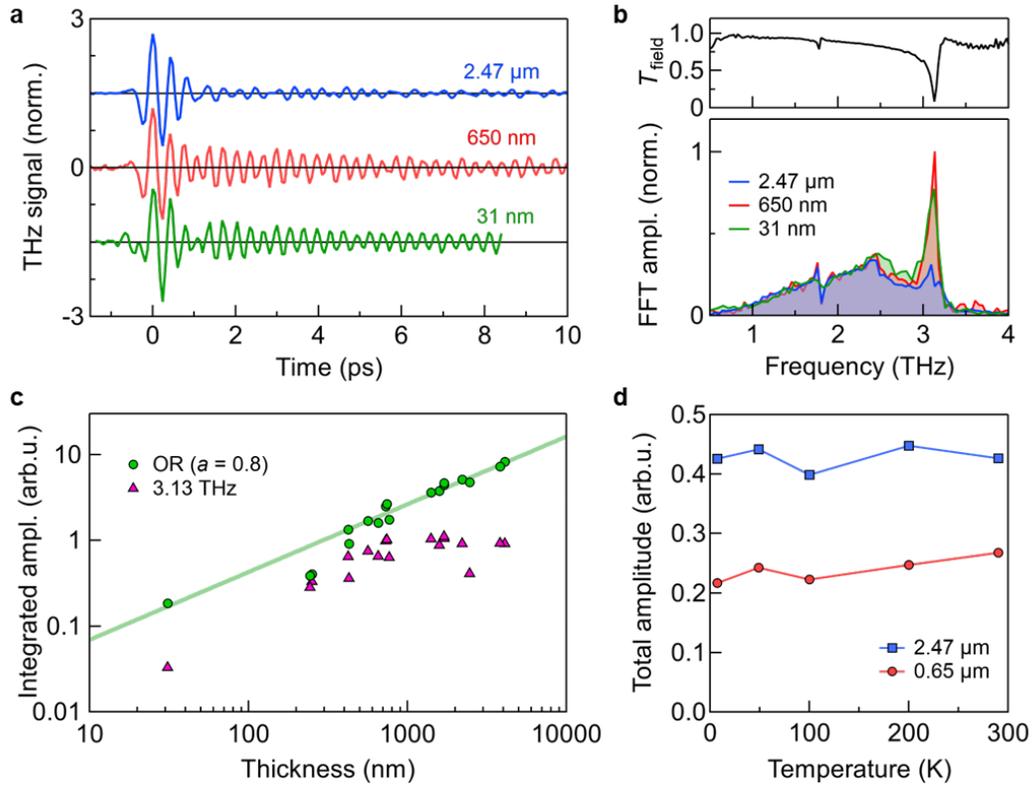

**Fig. 2 | Thickness dependence of THz emission.** (a) Normalized time-domain signal of emitted THz field for 2.47 μm (blue), 650 nm (red), and 31 nm (green) thick $NbOI_2$. (b) Lower panel shows the high-resolution frequency-domain spectra obtained from Fourier transform of the time-domain data shown in Fig. 2a. Upper panel shows a field transmittance spectrum determined in a separate THz-TDS experiment. A prominent, sharp THz emission peak emerges for the thin $NbOI_2$ flakes; this peak coincides with the TO phonon frequency at 3.13 THz (see the text). (c) Thickness dependences of the frequency-integrated THz emission amplitudes for the broadband optical rectification (green circles) and narrowband emission at 3.13 THz (purple triangles) obtained under the pump fluence of 1.24 mJ/cm$^2$. The solid line is the fit result to the power law dependence. (d) Temperature dependence of THz emission amplitude for different thicknesses, demonstrating that the THz emission does not change with temperature (see Supplementary Figs. S7 and S8 for time-domain traces).

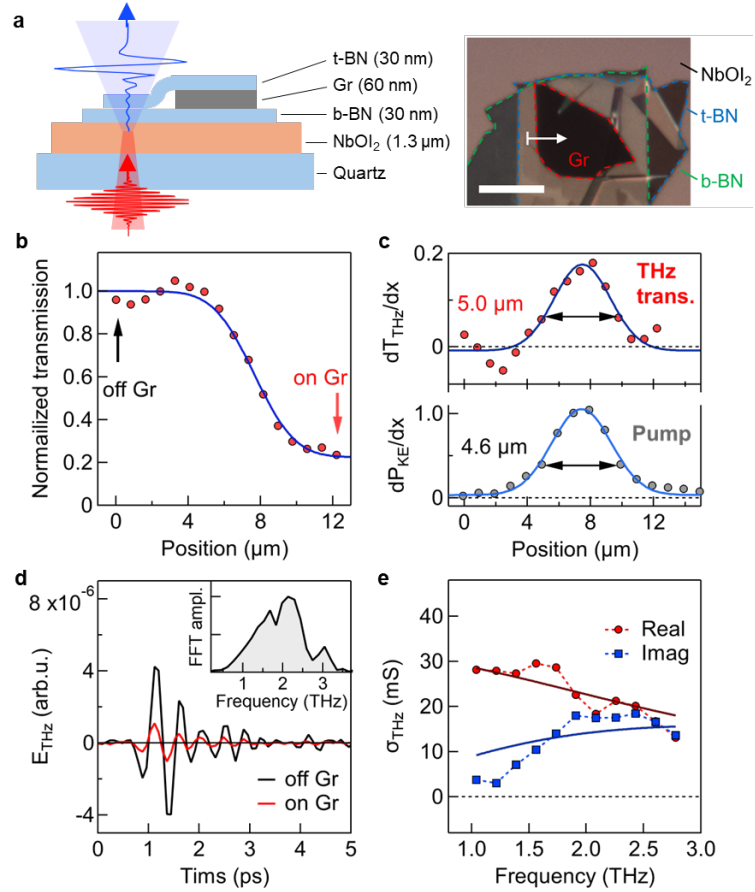

**Fig. 3 | *In-situ* and near-field THz-TDS in a vdW structure.** (a) Left panel shows a schematic side view of a vdW heterostructure for *in situ* THz spectroscopy. The heterostructure of *h*-BN/graphite/*h*-BN is prepared and transferred onto the THz emitter, which is a 1.3 μm thick flake of $NbOI_2$ on a 500 μm thick quartz substrate. Right panel shows an optical microscope image of the heterostructure. The scale bar is 20 μm. (b) The peak field strength of the transmitted THz light as a function of the sample position measured along the white arrow in the right panel of Fig. 3a. The transmittance is normalized to the value for the region without graphite. (c) Top panel shows the spatial derivative of the position-dependent THz transmittance in Fig. 3b. The blue curve is the Gaussian fit to the data, showing a diameter of 5.0 ± 0.8 μm for the THz field. The same, but position-integrated fit result is reproduced in (b) with a blue curve. The diameter is defined as the width of the Gaussian peak where the amplitude becomes 1/e of the central value. Lower panel shows the knife edge result for the 800-nm pump beam, showing a pump diameter of 4.6 μm at the sample position in terms of power. (d) Time-domain THz waveform taken on and off the graphite region, as indicated by the red and black arrows in Fig. 3b. The inset shows the Fourier-transformed amplitude of the waveform for the region off the graphite, i.e. the local reference THz spectrum, which covers up to 3.2 THz determined by 1-mm ZnTe EO detector. (e) THz complex conductivity spectrum of the graphite flake. The solid curves show fitting results to the Drude model (see main text).

**Methods**

Crystal growth of $NbOI_2$

Crystals of $NbOI_2$ were synthesized using a chemical vapor transport method. Niobium powder (0.1537 grams, Thermo Fischer Scientific, Puratronic 99.99% purity), $Nb_2O_5$ powder (0.1466 grams, Thermo Fischer Scientific, Puratronic 99.9985% purity), and $I_2$ chunks (0.7147 grams, Sigma Aldrich, 99.99+% purity) were loaded into a 12.7 mm o.d., 10.5 mm i.d. fused silica tube backfilled with argon. These reagent masses correspond to 1.0000 grams of $NbOI_2$ and 0.0150 grams of $I_2$ after these reagents have fully reacted. The excess iodine is the transport agent that facilitates crystal growth. The tube was flame sealed under ~30 mtorr of pressure to a length of 12 cm while the bottom of the tube was submerged under liquid $N_2$ to prevent volatilization of $I_2$. The tube was placed inside a computer-controlled, two-zone tube furnace and heated using the following heating profile where the source and sink side are the locations of the reagents and deposition areas of $NbOI_2$ crystals, respectively. Source side heating profile: Heat to 220°C in 3 hours, dwell for 24 hours, heat to 500°C in 24 hours, dwell for 24 hours, heat to 550°C in 6 hours, dwell for 144 hours, cool to ambient temperature in 6 hours. Sink side heating profile: Heat to 220°C in 3 hours, dwell for 24 hours, heat to 550°C in 24 hours, dwell for 24 hours, heat to 500°C in 6 hours, dwell for 144 hours, cool to ambient temperature in 6 hours. To remove excess iodine from the surface of the $NbOI_2$ crystals, the side of the tube with the crystals was heated to ~100°C and the other side was kept at ambient temperature to sublime the iodine away.

Preparation of $NbOI_2$ flakes

For the preparation of $NbOI_2$ flakes with thicknesses around 1 μm, we used thermal release tape (TRT). Bulk $NbOI_2$ was mounted to a glass microscope slide with double-sided Kapton polyimide tape (DSKT). The bulk crystal was cleaved using TRT (Semiconductor Equipment Corp. Revalpha RA-95LS(N)). The crystal on the TRT was thinned by several rounds of cleaving with Scotch tape, and the thickness was estimated by holding each flake against a light to check for translucency. Once the desired thickness was reached, each flake was placed on a z-cut quartz substrate (MTI Corporation). The sample assembly was heated to 135°C to remove the TRT, leaving the thin flake on the quartz substrate. The precise thickness was measured with a surface profilometer (Alpha Step D-600 KLA-Tencor). For high-resolution THz emission spectroscopy in Fig. 2b, we used a 2-mm thick z-cut quartz substrate. Otherwise, we used 0.5-mm thick z-cut quartz substrate.

For flakes thinner than around 500 nm, we used Scotch tape for tape exfoliation. Bulk $NbOI_2$ was mounted to Scotch tape and thinned by a few rounds of cleaving with Scotch tape. The $NbOI_2$ flakes on the tape were placed on a 0.5-mm thick fused silica substrate (Corning 7980) and heated to 70°C for 10 minutes. After the substrate was cooled down, the tape was removed from the substrate and the thin flakes were transferred to the fused silica substrate. The thickness of the thin flakes was determined by either Bruker Dimension FastScan AFM or Asylum Research Cypher S AFM.

Fabrication of vdW stack

The graphite and h-BN flakes were mechanically exfoliated from commercially purchased bulk graphite crystals from HQ Graphene. Prior to stacking, the graphite flake was characterized by an atomic force microscope (AFM) to measure the thickness with a Bruker Dimension Icon AFM. A standard polycarbonate based dry transfer technique was used to pick up the h-BN, graphite and h-BN flakes sequentially and melt down the stack on the $NbOI_2$ flake prepared by thermal release tape on a z-cut quartz substrate.

THz experiments

THz emission was measured using a home-built terahertz time-domain spectroscopy (THz-TDS) setup described in Ref. [55] with modifications. Fig. S10 depicts a schematic illustration. As a light source, we used either Ti:sapphire oscillator laser (Coherent, Vitara) or a Ti:sapphire regenerative amplifier (RA) (Coherent, Legend) seeded by the oscillator. The RA has a pulse duration of 30 fs, a repetition rate of 10 kHz, and a wavelength centered at 800 nm, while the oscillator has a repetition rate of 80 MHz. The results in Figs. 1 and 2 in the main text were obtained with the RA, where the output of RA was split into two beams for the pump and THz sampling. The pump beam was focused onto a sample on a quartz substrate, where the pump spot size at the sample position was determined to be around 50 μm (1/e radius) using a knife edge method. The pump light was incident from the quartz substrate side to characterize a true bandwidth of NbOI$_2$ without being affected by the THz absorption of quartz. The polarization of the pump beam was controlled using an achromatic half-wave plate (HWP) mounted on a motorized rotational stage. The THz radiation from the sample was collected using a parabolic mirror and passed through a high-resistivity Si wafer and a high-density polyethylene plate to block the optical pump, followed by a THz polarizer (PureWave) to define the polarization direction being measured. The THz light was then focused onto an electro-optic (EO) detector using another parabolic mirror along with an 800-nm sampling beam, by which the time-domain waveform of the THz electric field was detected via EO sampling. We used either 1-mm ZnTe or 0.2-mm GaP crystals for EO sampling. EO signal was recorded using a balanced detector and data acquisition card as described in Ref. [55]. The optical paths directing THz lights were enclosed and purged with dry air. The samples were placed in dry air at room temperature, except for temperature-dependent measurement, which was performed using a cryostat coupled to a closed cycle Helium recirculating system (Janis RGC4). Prior to all the THz experiments, the orientation of a NbOI$_2$ crystal placed in the THz setup was determined using visible polarimetry (described below) by exploiting its anisotropic optical transition (Fig. S3). We confirmed that quartz and fused silica substrates do not show THz emission under the fluence used. When we compare the THz emission property of NbOI$_2$ to that of ZnTe, we replaced the NbOI$_2$ sample with 200-μm ZnTe while keeping the experimental condition same.

The same setup was used to obtain the transmission spectrum reported Fig. 2b using the method described in Ref. [55]. Broadband THz probe light was generated with a two-color air plasma method from the RA output. This THz probe was focused onto an NbOI$_2$ flake mounted on a 1-mm precision pinhole (Thorlabs) that defines the sample area having high homogeneity. Reference TDS was taken with another blank precision pinhole. The transmittance was obtained by dividing the sample and reference data after Fourier transformation.

The THz microscopy results in Fig. 3 were obtained using the oscillator laser as the light source. A lens with a focal length of 1.5 cm was used to tightly focus the pump on the sample. The output from the balanced detector was recorded with a lock-in amplifier (Stanford Research System) by modulating the pump at 8.5 kHz with an optical chopper (Thorlabs). The sample was scanned using a motorized translational stage and controller (Newport).

The THz field amplitude was estimated based on the equation for the EO effect, $\Delta I/I_0 = 2\pi n_{\text{IR}}^3 r_{41} L E_{\text{THz}}/\lambda$ [56]. Here, $\Delta I$ is the THz-induced intensity difference measured by the balanced detector, $I_0$ the intensity of the incident sampling beam determined by blocking one of the detectors and using a calibrated neutral-density filter, $n_{\text{IR}}$ the near-infrared refractive index (2.8 for ZnTe at 800 nm), $r_{41}$ the electro-optic coefficient (3.9 x 10$^{-12}$ m/V for ZnTe), $L$ the crystal thickness (1 mm for ZnTe), $\lambda$ the wavelength of the sampling beam (800 nm), and $E_{\text{THz}}$ is the THz field strength. The THz signal reported in Fig. 1c in the main text is the measured $\Delta I/I_0$ using the 1-mm ZnTe.

Visible imaging and polarimetry

We built an external visible imaging and polarimetry system (see Fig. S10). We coupled it to the THz-TDS setup, which was used to remotely determine the crystal orientation on each flake prior to THz experiments, estimate the thickness from etalon interference, and verify the measured sample location and a good alignment to the EO sampling system. A white light (fiber-coupled illuminator, Thorlabs) was sent through a through hole on a parabolic mirror. The aromatic HWP described above was used to control the polarization of the incident white light. The 1.5-cm focal length lens was used to focus the white light and collect the reflected light. When imaging, an additional lens was inserted to make a Köhler-type illumination. The collected reflected light was sent to an imaging camera or a fiber end with a resettable mirror. The other end of the fiber was coupled to a liquid-nitrogen-cooled CCD camera equipped with a monochromator (Princeton), which measures the polarization dependent reflectivity.

**Acknowledgements**

Experiments on the discovery of THz emission (figure 1) and implementation in near-field microscopy and spectroscopy (figure 3), as well as sample synthesis and characterization, were supported by the Materials Science and Engineering Research Center (MRSEC) through NSF grant DMR-2011738. Experiments on thickness dependence and THz emission mechanisms (figure 2) were supported by the US Army Research Office, grant number W911NF-23-1-0056. C.Y.H. was supported by the Taiwan-Columbia Fellowship funded by the Ministry of Education of Taiwan and Columbia University. F.S. acknowledges support from the Center on Programmable Quantum Materials, an Energy Frontier Research Center funded by the U.S. Department of Energy (DOE), under award number DE-SC0019443. XYZ acknowledges support by the Max Planck – New York City Center for Non-Equilibrium Quantum Phenomena for fruitful collaborations, interactions and discussions with the Max Planck Institutes (MPIs), particularly Mischa Bonn of MPI-Mainz, and Michael Fechner and Hope M. Bretscher of MPI-Hamburg. T.H. acknowledges support by JSPS Overseas Postdoctoral Research Fellowship program.

**Author contributions**

T.H. and X.Y.Z. conceived this work. T.H. carried out optical measurements. C.Y.H., N.O., and F.S. prepared the exfoliated samples. Y.L. and D.D.X. fabricated the vdW heterostructures. D.G.C. synthesized the crystals under the supervision of X.R. X.Y.Z. supervised the project. The manuscript was prepared by T.H. and X.Y.Z. in consultation with all other authors. All authors read and commented on the manuscript.

**Competing Interests**

The authors declare that they have no competing interests.

**Supplementary Information for**

**A 2D van der Waals Material for Terahertz Emission with Giant Optical Rectification**

Taketo Handa[1], Chun-Ying Huang[1], Yiliu Li[1], Nicholas Olsen[1], Daniel G. Chica[1], David D. Xu[1], Felix Sturm[2], James W. McIver[2], Xavier Roy[1], and Xiaoyang Zhu[1,*]

1. Department of Chemistry, Columbia University, New York, NY 10027, USA
2. Department of Physics, Columbia University, New York, NY 10027, USA

[*]Corresponding author. E-mail: xyzhu@columbia.edu

**Supplementary Text**

Pump fluence dependence for optical rectification

Non-centrosymmetric materials with broken inversion symmetry possesses nonzero second-order nonlinear susceptibility $\chi^{(2)}$, which can give rise to SHG and optical rectification in response to an incident electric field. The induced second-order nonlinear polarization $P^{(2)} = \epsilon_0 \chi^{(2)} [E_i(t)]^2$ can be a source of electromagnetic radiation. The induced second-order nonlinear polarization reads $P^{(2)} = \epsilon_0 \chi^{(2)} [E_i(t)]^2 = \epsilon_0 \chi^{(2)} E_0^2 \cos^2 \omega_0 t = E_0^2 (\cos 2\omega_0 t + 1)/2$ for an oscillating electric field $E_i(t) = E_0 \cos \omega_0 t$. The second term in the right-hand side that is not a function of time describes the optical rectification. The generated THz field is linearly proportional to $P^{(2)}$ and thus is proportional to $E_0^2 = I_0$, where $I_0$ is the pump fluence. Here, for simplicity we considered scalar values here.

High saturation fluence of NbOI$_2$

The high saturation threshold for THz generation of NbOI$_2$ likely originates from its wide bandgap along the polar *b*-axis. The exciton-like absorption peak in the linear absorption spectrum is located around $E_0$ = 3.5 eV (354 nm)[1]. With the pump energy *hv* of 1.55 eV (800 nm) used in the present study, $2hv < E_0$, and thus there is no real states available that can induce the two-photon absorption process. In other words, the third-order nonlinearity is relatively small at least with the current pump wavelength. On the other hand, the bandgap of ZnTe is narrower, around $E_{0,\text{ZnTe}}$ = 2.3 eV. Because $2hv > E_{0,\text{ZnTe}}$, the 800-nm pump can induce enhanced two-photon absorption due to resonance effects. Under higher fluences, the two-photon absorption process competes with optical rectification, which reduces the net efficiency of optical rectification and leads to the saturating trend for the fluences higher than 1 mJ/cm$^2$ for ZnTe.

Meanwhile, the second-order nonlinearity of NbOI$_2$ is considered to be significantly enhanced along the polar axis due to the ferroelectric polarization[1]. The resulting large second-order nonlinearity gives rise to very efficient optical rectification, while the wide bandgap of NbOI$_2$ contributes to the high saturation fluence and excellent linearity for the wide range of fluence.

Narrowband THz radiation at 3.13 THz

We have observed the persistent coherent THz radiation, peaked at 3.13 THz and featuring a very narrow bandwidth, only when the flake thickness is smaller than ~1 μm (Fig. 2). We suggest here that this narrowband emission is due to oscillating macroscopic ferroelectric dipoles caused by the ultrafast launching of coherent soft transverse optical (TO) phonons.

Fig. 2b (top panel) shows the THz field transmittance for a freestanding NbOI$_2$ (2.6 μm) measured with an incident field along the polar *b*-axis. It reveals the presence of TO phonon at 3.13 THz, which matches the prominent peak in the THz emission spectrum. Our recent density functional theory (DFT) calculations revealed the nature of this TO mode. It corresponds to the motion of Nb and O atoms along the crystallographic *b*-axis, oscillating in-phase[2]. This motion changes the relative length of O-Nb-O bonds, modulating the magnitude of ferroelectric spontaneous polarization (see Fig. 1b in the main text), reminiscent of soft TO modes in displacive-type ferroelectrics. This theoretical insight of the soft TO feature of the 3.13 THz mode was further corroborated by the absence of the 3.13-THz TO phonon when the probe THz field is along the non-polar *c*-axis (see Fig. S11).

Next, we discuss the launching of coherent wavepacket of this TO phonon. We recently studied coherent phonon properties of NbOI$_2$ by performing visible transient reflectivity[2]. This experiment revealed that ultrafast optical excitation (above gap) launches prominent, long-lived coherent phonons with a frequency of 3.13 THz. The frequency and the long coherence time of the coherent phonons in transient reflectivity match the observed coherent narrowband THz emission at 3.13 THz in Fig. 2b. Although the pump wavelength of 800 nm for the present THz emission experiment is below the gap, we consider that the 800-nm pump also launches the same coherent phonons via the impulsive stimulated Raman scattering mechanism. Indeed, a strong signal of the 3.13 THz mode was verified in a ground-state Raman scattering spectrum using 633 nm as a fundamental (below gap). Note that the 3.13-THz mode is both Raman and IR active due to the low *C*2 symmetry of NbOI$_2$.

Based on these experimental and theoretical results, we conclude that the ultrafast near-infrared pump launches coherent soft TO phonons at 3.13 THz that induce the time variation of the spontaneous polarization with a frequency of 3.13 THz, and these coherently oscillating dipoles can be a macroscopic and dipole-like radiation source with a monochromatic frequency. We stress that this mechanism is possible because spontaneous polarizations and their modulation by coherent phonons are both macroscopic quantities, enabling the direct coupling to transverse electromagnetic waves. In contrast, conventional coherent TO phonons (not ferroelectric modes) cannot be such a radiation source because they are not accompanied by a macroscopic field[3]. This scenario can be seen in early works[3–5] studying THz emission from conventional non-ferroelectric semiconductors, where no peak structure at TO frequency has been observed. Despite a seemingly straightforward mechanism, THz radiation from soft TO phonons has been very scarce. In supplementary material in Ref. [6], a signal associated with soft phonons from Sn$_2$P$_2$S$_6$ was suggested, but the signal was weak and buried in an optical rectification signal without careful spectral and polarization analysis. Also, the observation was not associated with oscillating macroscopic dipoles but rather with nonlinear polarization.

The narrow THz radiation at 3.13 THz was observed for reduced thickness. One possible origin for this thickness dependence is the re-absorption of the emitted radiation by the same TO mode. Another possible origin is intrinsic coupling between soft TO phonons and THz electromagnetic mode. Recently in Ref. [7], it was suggested that such a coupling can strongly renormalize the dynamics of soft phonons, in particular effectively acting as an additional damping. As a consequence, THz radiation from soft TO phonons was predicted only for thin limits, thinner than a few hundred of nm in the case of BaTiO$_3$ with in-plane polarization. Although we are not sure which of these mechanisms dominantly govern the observed thickness dependence, the 2D nature and the intrinsic in-plane ferroelectricity of NbOI$_2$ will be ideal to systematically study the unconventional narrowband THz radiation from the soft TO mode in future studies.

We note that we can exclude another possible origin of phase matching. A sharp THz radiation peak was observed from Bi$_4$Ge$_3$O$_{12}$[8]. It was explained by an enhanced phase matching condition and longer coherence length, which exist only for a narrow frequency region. However, the Bi$_4$Ge$_3$O$_{12}$ crystal used in Ref. [8] was 1 mm, where the phase matching condition is crucial. Meanwhile, the thicknesses studied in the present work were around 1 µm. In this thin region, the phase matching condition is always met due to the long wavelength of THz radiation, 96 µm at 3.13 THz.

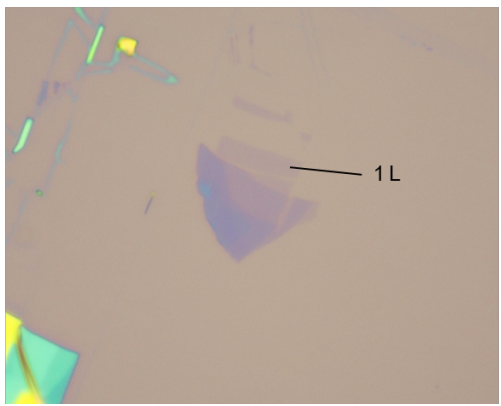 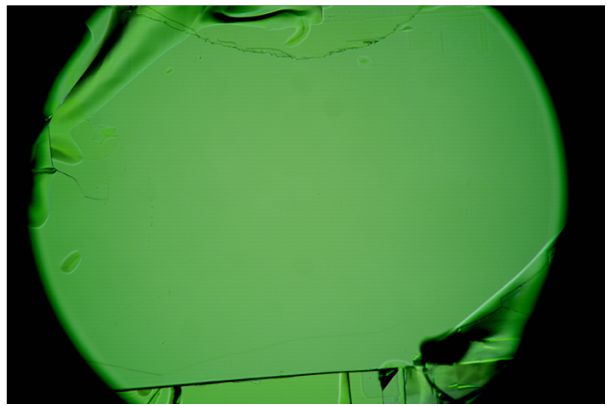

**Fig. S1.** (a) NbOI$_2$ monolayer obtained by a Scotch tape exfoliation and (b) 0.565-μm thick NbOI$_2$ flake obtained by a thermal-release tape exfoliation.

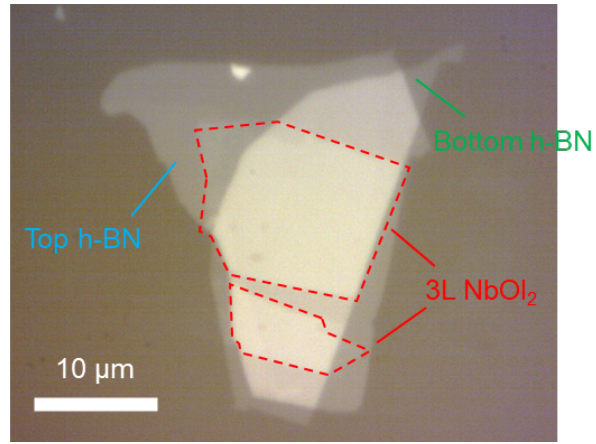

**Fig. S2.** A vdW structure of h-BN/tri-layer $NbOI_2$/h-BN prepared by a conventional polycarbonate based dry transfer technique.

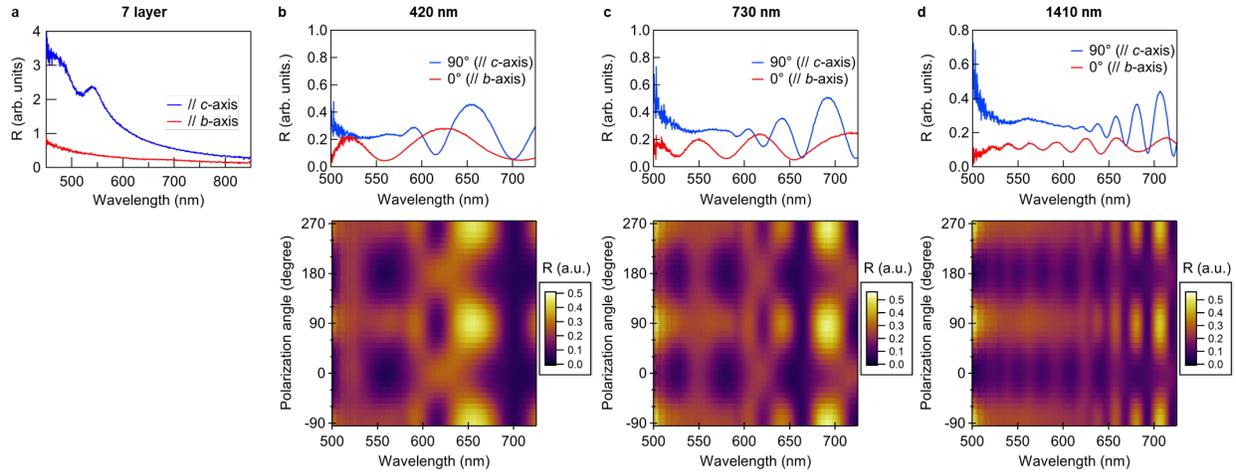

**Fig. S3.** (a) Reflectance spectra of NbOI$_2$ (7 layer) obtained with the incident light polarization along the polar *b*-axis (red) and the non-polar *c*-axis (blue). Data is taken from Ref. [2]. Along the polar *b*-axis, the direct excitonic bandgap is wide (higher than 2.8 eV), so the pump wavelength of 800 nm used in the present study for THz emission does not induce photocarrier generation via linear absorption. (b-d) Reflectance spectra for thicker flakes measured inside the THz emission setup. While lower panels show the full polarization dependence, the top panels show the spectra along the polar *b*-axis (red) and the non-polar *c*-axis (blue). The thicknesses are indicated at the top of each figure. A clear Fabry-Perot interference can be seen in the reflectance spectrum along the polar *b*-axis for all the wavelengths measured (red curves), verifying the transparency in the visible to near-infrared region and the high surface quality that can induce the Fabry-Perot interference. When the light is polarized along the non-polar *c*-axis, a peak structure can be seen at 560 nm due to the excitonic transition, and a Fabry-Perot interference starts for the longer wavelengths. These polarization dependent reflectance data were used (i) for the determination of crystal orientation at the focus position in the THz setup and (ii) for the optical estimation of the thickness. Specifically, at the wavelength around 500 nm, the reflectance is always higher when the incident polarization is along the non-polar *c*-axis than being along the polar *b*-axis, which allows the unambiguous in-situ optical determination of the crystal orientation.

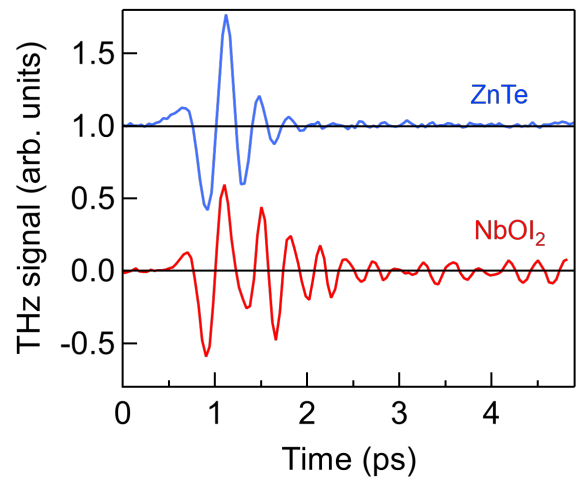

**Fig. S4.** Time-domain terahertz emission signal from 200-μm ZnTe (top) and 2.5-μm NbOI$_2$ (bottom) recorded using a 0.2-mm GaP as an EO detector.

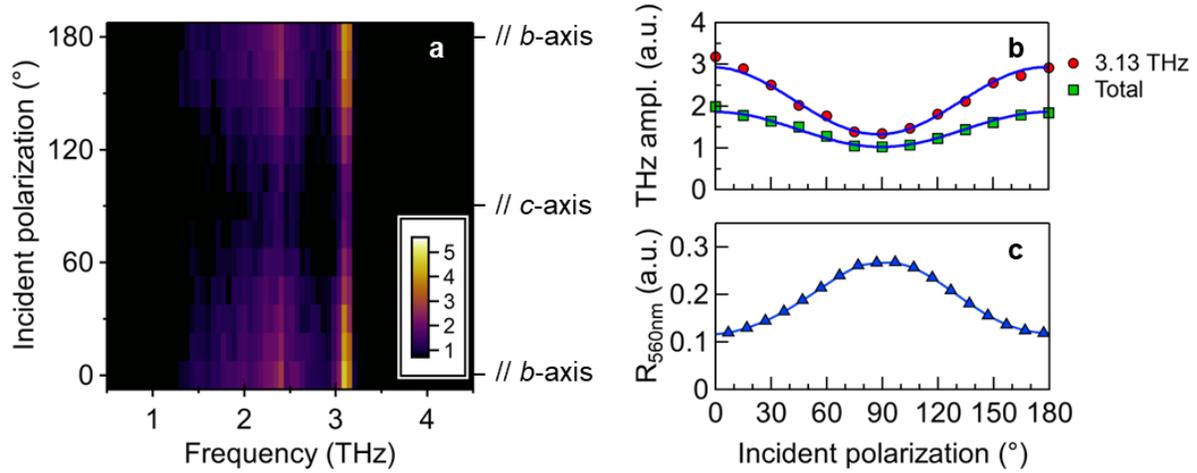

**Fig. S5.** (a) Pump polarization dependence of THz emission spectrum for 650 nm thick $NbOI_2$ at room temperature. The incident angle of 0° corresponds to the polarization of pump 800 nm laser parallel to the polar *b*-axis. For the EO detection, EO crystal orientation and gate polarization were chosen to measure the polarization component of emitted THz radiation parallel to the *b*-axis. (b) spectral amplitudes for the 3.13 THz peak (red dots) and the total amplitude (green squares), showing the stronger amplitudes when the pump polarization is parallel to *b*-axis. (c) Incident light polarization dependence of the reflectance measured at 560 nm of the same flake. Due to the anisotropic optical property discussed in Fig. S3, one can use the anisotropic reflectance to determine the orientation, and subsequently one can determine the pump polarization to obtain higher THz emission.

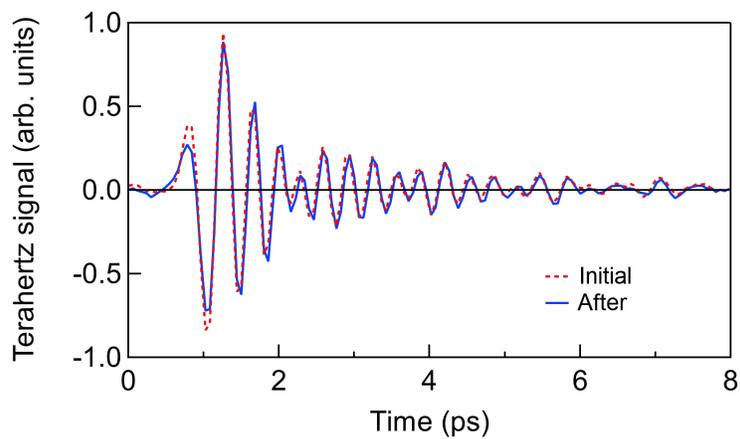

**Fig. S6.** The comparison of THz emission signals from 3.8 μm thick NbOI$_2$ that was freshly cleaved (dotted red) and after the laser irradiation with 1.24 mJ/cm$^2$ for 20 hours in dry air (solid blue). This comparison shows a negligible change of the emission property.

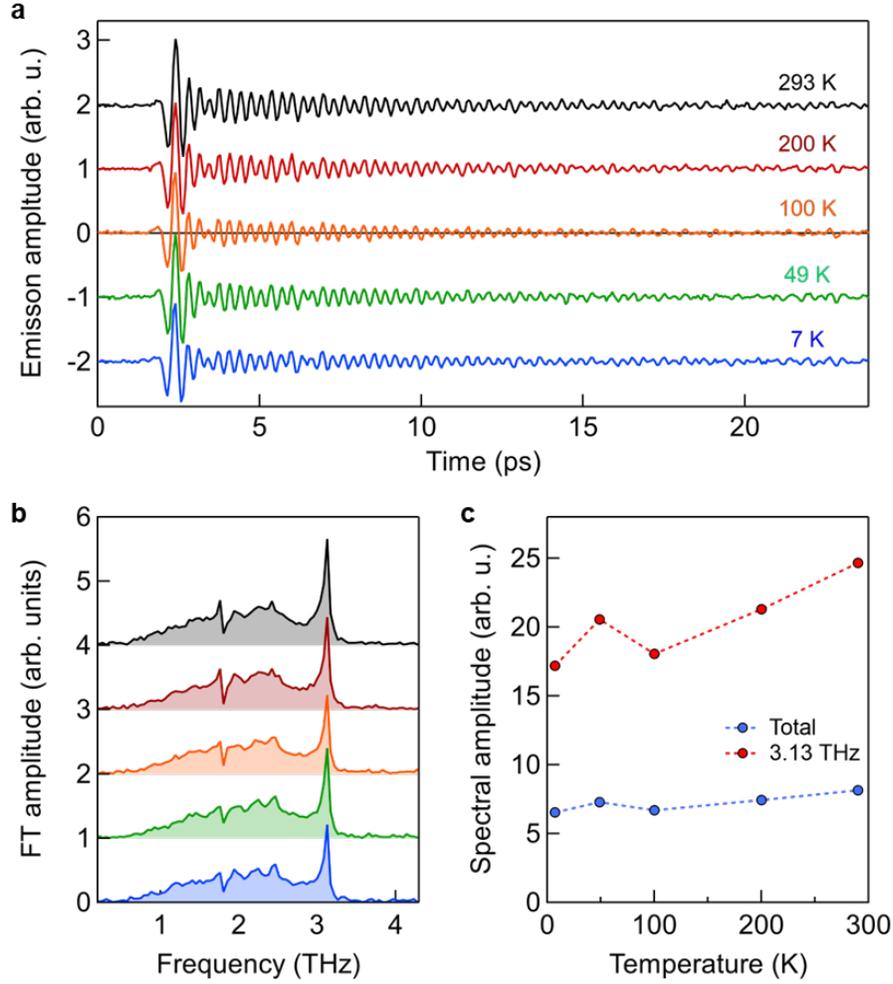

**Fig. S7.** Temperature dependence of THz emission properties for 650 nm thick $NbOI_2$. (a) The time-domain waveforms, (b) frequency-domain spectra obtained by Fourier transformation for the entire time window, and (c) the spectral amplitudes for the entire frequency from 0.26 to 3.4 THz (red) and for the 3.13 THz peak (blue). We observed almost temperature-independent THz emission properties. The position of the sharp emission peak does not change.

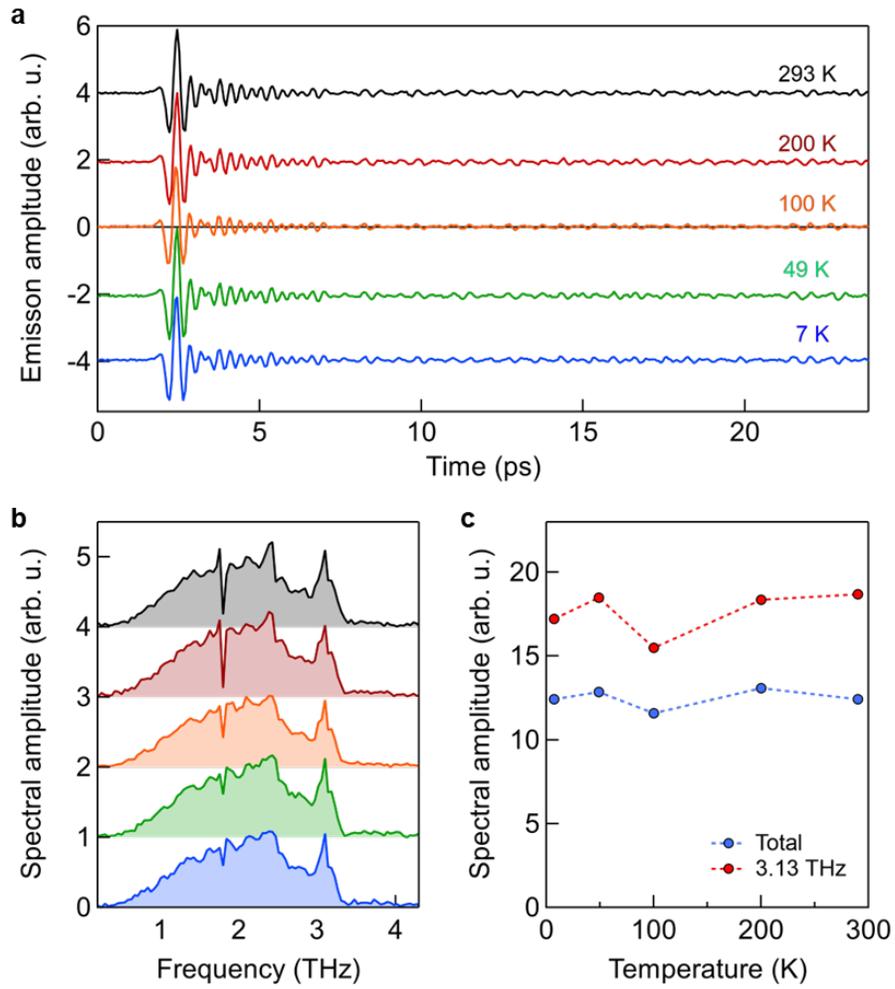

**Fig. S8.** Same plot as Fig. S7, but for 2.47 μm thick $NbOI_2$, showing the temperature-independent THz emission properties also for the thicker flake.

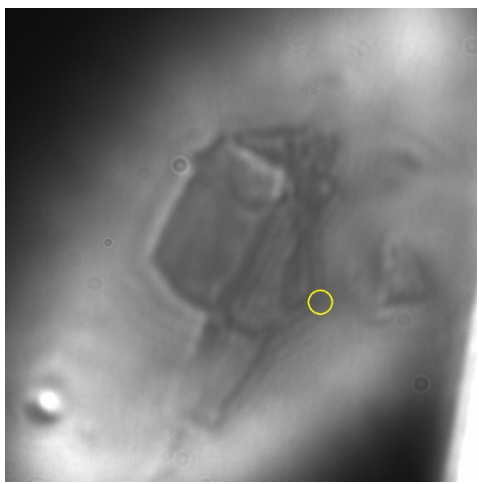

**Fig. S9.** Optical image of the vdW stack in Fig. 3, seen through from the 1.3-µm NbOI$_2$ side at the sample position in the THz setup. Because of the reasonable transparency of NbOI$_2$, the vdW stack can be observed and carefully aligned with the pump laser spot. This image was taken with the resettable 1.5-cm focal length lens without introducing any modification to the THz-TDS setup (see Fig. S10). The image is flipped to allow a direct comparison to the image in Fig. 3a. The yellow circle was used for an alignment purpose.

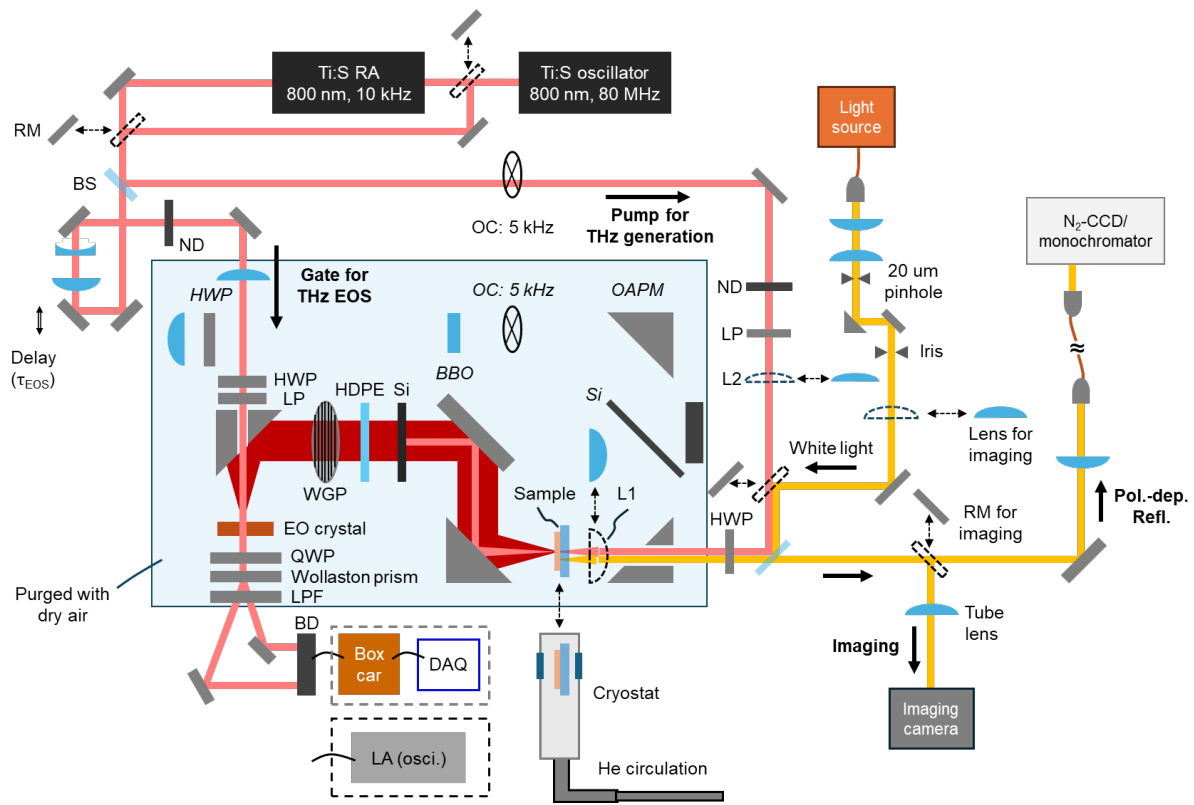

**Fig. S10.** Schematic illustration of the THz emission setup coupled with visible imaging and polarimetry. RA, regenerative amplifier; RM, resettable mirror; BS, beam splitter; ND, neutral-density filter; HWP, half-wave plate; QWP, quarter-wave plate; LP, linear polarizer; OAPM, off-axis parabolic mirror; WGP, wire grid polarizer; BBO, beta barium borate; OC, optical chopper; LPF, longpass filter; BD, balanced detector; DAQ, data acquisition card; LA, lock-in amplifier (used for oscillator-based experiments). L1 (L2) is a singlet lens with a focal length of 1.5 cm (30 cm) used for oscillator- (RA-) based experiments. The optics described in the italic are used for THz-TDS, see Ref. [9] for details.

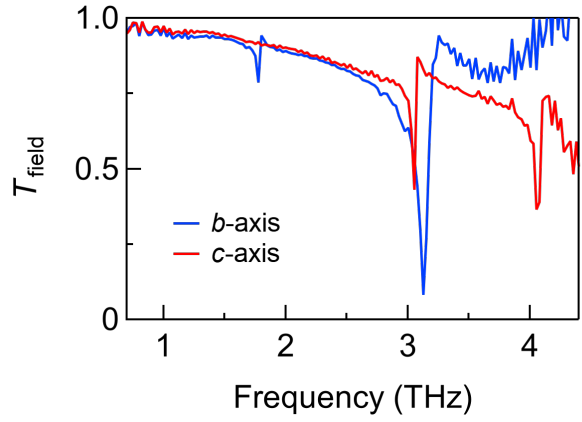

**Fig. S11.** THz field transmittances of 2.6 µm thick free-standing $NbOI_2$ along the crystallographic *b*-axis (red) and *c*-axis (blue). This polarization resolved transmittance reveals the existence of the polar TO phonon mode at 3.13 THz (elucidated also by DFT calculations) along *b*-axis, but not along *c*-axis.

## References

1. Abdelwahab, I. *et al.* Giant second-harmonic generation in ferroelectric NbOI2. *Nat. Photonics* **16**, 644–650 (2022).
2. Huang, C.-Y. Carrier Coupling to Ferroelectric Order in a 2D Semiconductor. *arXiv* (2024).
3. Dekorsy, T., Auer, H., Bakker, H. J., Roskos, H. G. & Kurz, H. THz electromagnetic emission by coherent infrared-active phonons. *Phys. Rev. B* **53**, 4005–4014 (1996).
4. Dekorsy, T. *et al.* Emission of Submillimeter Electromagnetic Waves by Coherent Phonons. *Phys. Rev. Lett.* **74**, 738–741 (1995).
5. Tani, M. *et al.* Terahertz radiation from coherent phonons excited in semiconductors. *J. Appl. Phys.* **83**, 2473–2477 (1998).
6. Sotome, M. *et al.* Ultrafast spectroscopy of shift-current in ferroelectric semiconductor Sn2P2S6. *Appl. Phys. Lett.* **114**, 151101 (2019).
7. Zhuang, S. & Hu, J.-M. Role of polarization-photon coupling in ultrafast terahertz excitation of ferroelectrics. *Phys. Rev. B* **106**, L140302 (2022).
8. Takeda, R., Kida, N., Sotome, M., Matsui, Y. & Okamoto, H. Circularly polarized narrowband terahertz radiation from a eulytite oxide by a pair of femtosecond laser pulses. *Phys. Rev. A* **89**, 033832 (2014).
9. Handa, T. *et al.* Spontaneous exciton dissociation in transition metal dichalcogenide monolayers. *Sci. Adv.* **10**, eadj4060 (2024).